# Analyzing Web Behavior in Indoor Retail Spaces

Yongli Ren
School of Computer Science and Information Technology, RMIT University, Australia.
Email: yongli.ren@rmit.edu.au

Martin Tomko
Department of Geography, University of Zurich, Zurich, Switzerland.
Email: martin.tomko@geo.uzh.ch

Flora Salim and Kevin Ong and Mark Sanderson
School of Computer Science and Information Technology, RMIT University, Australia.
Email: {flora.salim, kevin.ong, mark.sanderson}@rmit.edu.au

**Abstract** We analyze 18 million rows of Wi-Fi access logs collected over a one year period from over 120,000 anonymized users at an inner-city shopping mall. The anonymized dataset gathered from an opt-in system provides users' approximate physical location, as well as Web browsing and some search history. Such data provides a unique opportunity to analyze the interaction between people's behavior in physical retail spaces and their Web behavior, serving as a proxy to their information needs. We find: (1) the use of Wi-Fi network maps the opening hours of the mall; (2) there is a weekly periodicity in users' visits to the mall; (3) around 60% of registered Wi-Fi users actively browse the Web and around 10% of them use Wi-Fi for accessing Web search engines; (4) people are likely to spend a relatively constant amount of time browsing the Web while their visiting duration may vary; (5) people tend to visit similar mall locations and Web content during their repeated visits to the mall; (6) the physical spatial context has a small but significant influence on the Web content that indoor users browse; (7) accompanying users tend to access resources from the same Web domains.

## Introduction

While the use of the Web is well understood in many contexts, there is a new context emerging which is little understood: Web access in large indoor spaces, such as shopping malls, airports, universities, and museums. Indoor retail spaces impose various physical, social, and technical constraints, such as location, layout, opening hours, and Wi-Fi connectivity. For example, shopping malls normally open in the morning and the shops in the malls close in the evening while cinemas and restaurants may open until later. Shoppers are constrained by the time available for making purchases. However, those constraints are often implicitly expressed. Furthermore, owners of these spaces design and manage them under certain economic rationale (Vernor, Amundson, Johnson, & Rabianski, 2009), e.g. the principle of cumulative attraction where similar retail shops tend to be placed near each other. Market management research demonstrates that the social context of retail shopping has influences on customers' shopping behavior (Evans, Christiansen, & Gill, 1996; Khare, 2012).

In many indoor spaces, free Wi-Fi is increasingly available. This creates an environment where shoppers can remain connected without worrying about the costs of their data usage. Visitors are thus exposed to an engineered environment with a mix of physical, social, and technical factors influencing their needs and desires. Understanding users' physical and Web behavior is fundamental to improving the designs of indoor services – both the physical retail services and the accompanying Web services.

Corresponding Author: Yongli Ren, School of Computer Science and Information Technology, Melbourne, VIC 3001, RMIT University, Australia. Email: yongli.ren@rmit.edu.au

Previous research focused on either indoor spaces (Biczok, Martinez, Jelle, & Krogstie, 2014), general Web browsing/searching (Spink, Jansen, Wolfram, & Saracevic, 2002; Silverstein, Marais, Henzinger, & Moricz, 1999; Jansen, Ciamacca, & Spink, 2008), or general mobile Web browsing/searching (Cui & Roto, 2008; Church, Smyth, Cotter, & Bradley, 2007; Church & Smyth, 2009; Nylander, Lundquist, & Brännström, 2009; Kamvar & Baluja, 2006), but rarely in connection. This study tries to fill this gap by investigating the *Web activities* of visitors of an inner-city shopping mall in Sydney, Australia, together with their corresponding *physical activity* within the mall. In this paper, Web activities are analyzed based on a large-scale log of Web activity of around 120,000 users, collected over a 1 year period. Additional data about the *physical environment* are provided by the owner of the mall, including the floor maps of the stores, their shop categories, and the location of the Wi-Fi access points.

The diverse aspects of the physical and Web behavior of indoor users and their relationships are explored through the following research questions:
- Does the use of Wi-Fi network correspond to the opening hours of the mall?
- Do users tend to visit the retail mall on a certain frequency?
- Are users likely to access the Web while visiting the mall?
- Do users always keep accessing the Web during their visits?
- Do users tend to visit similar mall locations and Web content during their repeated visits to the mall?
- Does users' Web behavior correlate with the indoor spatial context?
- Does users' social context correlate with their Web behavior?

The main contribution of this paper is
- a comprehensive report of user indoor behavior;
- an analysis of the correlation between users' physical visiting patterns and their Web behaviors;
- the establishment of the significant influence of the physical spatial context on the content that indoor users consume on the Web, and for the first time, this is done on a much more detailed level of Wi-Fi APs and shop categories;
- and finally, the analysis of the correspondence between indoor users' social context and their Web behaviors.

To the best of our knowledge, this is the first such research conducted on a dataset of a significant size in large indoor spaces.

After the review of related work, we describe the collected data, and define terminology. Then, we explore the basic physical behavior and Web behavior for indoor visitors to retail environments and provide the analysis of the patterns in indoor physical and Web behavior; We interpret the findings and propose avenues for future work, and finally conclude the paper.

# Related Work

Information behavior is a term to describe the ways in which people interact with information (Bates, 2010). When designing information services for (indoor) mobile use, one should consider the purpose for which mobile devices are used. Here we review users' information behavior on the Web, where users either *search* for information in a goal oriented manner or *browse* the content to satisfy their information needs. The use of Web usage patterns as a means to personalize content served to users is an idea explored extensively over the last 15 years (for an early overview, see (Mobasher, Cooley, & Srivastava, 2000)). Web usage mining as a way to infer individualized content has been perceived superior to manually created profiles or individual user content rating-based recommendations due to the reduced subjectivity of the method, relying on actual activity patterns (Mobasher et al., 2000). The connection between indoor physical behavior (captured using mobile devices) and Web behavior has so far been insufficiently investigated – in particular on large-scale real-world datasets.

*Information behavior and the Web*
Web search has been widely studied from many aspects. Two early studies of desktop based Web search used logs from Excite (Jansen, 2000; Spink, Wolfram, Jansen, & Saracevic, 2001; Wolfram, Spink, Jansen, & Saracevic, 2001; Spink et al., 2002) and AltaVista (Silverstein et al., 1999). They examined key characteristics of Web search queries, such as the number and distribution of terms, the number and distribution of queries within sessions, and the topical search categories. Web use has, however, changed significantly in the recent years. Some more recent Web search studies focused on other perspectives, e.g. specific searches on the Web (Jansen et al., 2008), comparison of search in different IR environments (Wolfram, 2008), queries of children (Duarte Torres, Hiemstra, & Serdyukov, 2010), geographic queries (Sanderson & Kohler, 2004; Gan, Attenberg, Markowetz, & Suel, 2008; Aloteibi & Sanderson, 2014), religious information in search engines (Wanchik, Clough, & Sanderson, 2013), temporal characteristics of query topical categories (Beitzel, Jensen, Chowdhury, Grossman, & Frieder, 2004) and sponsored search (Pandey, Punera, Fontoura, & Josifovski, 2010). There are some other recent studies focusing on analysis of Web logs. For example, Kumar and Tomkins (2010) studied the characterization of general online browsing behavior by proposing a pageview taxonomy: content, communication and search, and found that approximately half of the pageviews are *content*, one-third are *communication* and the rest are *search*. West, White, and Horvitz (2013) studied the spatiotemporal characteristics of population-wide dietary preferences by analyzing a large Web log collected via a Web-browser add-on. Specifically, they applied the number of recipes that users searched as a proxy for their food consumption, and they found that there were two periodic components in users' dietary preferences, one yearly and the other weekly, and the regional differences were also discovered.

*Mobile Web use*
Mobile Web use is significantly different from desktop (Kamvar & Baluja, 2006), e.g. how, when and where users search and browse the Web. Cui and Roto (2008) presented a study on how people use the Web on mobile devices, focusing on contextual factors and Web activities. They found people tend to use mobile Web in a stationary environment and in short sessions, and proposed a Web activity taxonomy: *information seeking*, *communication*, *transaction*, and *personal space extension*. Church and Smyth (2007) focused on the differences between mobile browsing and mobile searching, showing that mobile browsing was more common than mobile searching, although the latter was increasingly popular. Their follow-up work (Church & Smyth, 2009) analyzed the intent behind mobile information needs through a diary study. They found mobile needs differ significantly from general Web needs, as users are normally on-the-move. Thus, context influences the types of information, the goal, and the topics that users are interested in.

*Contextual influence on Web use*
Other studies investigated the contextual influence on mobile Web use. An extensive study by Lee, Kim, and Kim (2005) resulted in the proposal of a classification of contexts of mobile Web usage. The researchers showed that mobile Web use was skewed towards a handful of popular contexts (e.g. ringtone download, news, or weather services). Sohn, Li, Griswold, and Hollan (2008) from a diary study found that 72% of the participants' mobile information needs were prompted by contextual factors. Hinze, Chang, and Nichols (2010), using a small-scale diary study, reported a significant impact of contextual factors, e.g. location, conversation, activity, etc. They found that the identification of places can help to infer context but that use of query keywords could not be used to establish context. Teevan, Karlson, Amini, Brush, and Krumm (2011) performed a similar study on a larger scale, finding that mobile local searches were strongly influenced by context (e.g, geographic features, temporal aspects, and searchers' social context). Chua, Balkunje, and Goh (2011) examined what contextual factors triggered mobile information needs and what influenced those needs. They revealed that location, intended activity, and social surroundings *triggered* information needs while location, time, current activity, and social surrounding *influenced* information needs. Finally, Church and Oliver (2011) noticed how users increasingly use mobile Internet in more stationary and familiar settings and explored the popularity of mobile usage in different contexts.

All this previous work only modeled spatio-temporal context coarsely, e.g., "at home/work", "traveling abroad", "with friends/family", "in transit/commuting". In our study, the collected data provides a chance to investigate indoor users' Web behaviors at a detailed level.

*Indoor behavior tracking*
Indoor movement is structured by hallways and rooms (Jensen, Lu, & Yang, 2010), segregating spaces hierarchically by functional, organizational and social constraints (Richter, Winter, & Santosa, 2011). The structure of indoor space has been extensively analyzed by researchers of indoor navigation systems (Ruetschi, 2007; Richter et al., 2011), and related to the constraints the space imposes on movement. The structure of a space and the arrangement of signal beacons used for positioning have been used to improve the accuracy of indoor positioning (Bai et al., 2014; Bell, Jung, & Krishnakumar, 2010). Biczok et. al. (2014) analyzed users' indoor spatial mobility through MazeMap, a live indoor/outdoor positioning and navigation system. They found strong logical ties between different locations in users' spatial mobility. A model to classify indoor trajectories based on cellular spaces was proposed in (Kang, Kim, & Li, 2009) and contributes an important method to similarity analysis of coarsely expressed trajectories.

The LiveLabs project (Misra & Balan, 2013) is an example of an in-device positioning approach for indoor user behavior tracking using a smartphone app to track users indoors. Part of this project is a controlled study of thirty participants in a shopping mall to infer the buying intent of shoppers (Sen et al., 2014).

Since the organizational requirements of indoor positioning are poorly understood (Kjærgaard et al., 2014), most work focuses on limited populations of individuals over limited periods of time in instrumented settings. In contrast (as in our study), most indoor environments are set up with Wi-Fi networks to primarily provide Internet access to visitors and are optimized for coverage rather than positioning accuracy. The utility of large scale indoor tracking datasets collected as a by-product of their primary purpose over a long time for user behavior analysis is thus unknown and the applicability of insights from experiments conducted in carefully instrumented environments is uncertain.

# Data Acquisition and Processing

We study an anonymized dataset of Internet accesses by registered users of a free opt-in Wi-Fi network operated by a large inner-city shopping mall covered by 67 Wi-Fi Access Points (AP) across 90,000 square meters. The mall contains over 200 stores, and they belong to 34 shop categories as defined by the mall operator; sample shop categories are shown in Table 1. Floor plans of the mall were overlaid with AP locations and the service areas of the APs were approximated by Voronoi regions (Okabe, Boots, Sugihara, & Chiu, 1999), each centered on a single AP, that encompass all the points that are closest to that AP. The regions were manually rectified to correspond better with the frontages of physical stores in the mall (Fig. 1). Shop frontages are the main determinants of context as the Wi-Fi network is meant to cover common spaces in the mall.

Table 1. Sample shop categories

| Category | Category |
|---|---|
| Women's Fashion | Men's Fashion |
| Fine Jewellery | Music/Videos/DVDs |
| Sport | Toys & Hobbies |
| Kitchenware/Tableware | Home Décor |
| Computer Hardware & Software | Liquor |
| Furniture/Floor Coverings | Hair & Beauty |
| Fruit & Vegetable | Groceries |
| Restaurant | Cinemas |
| Travel | Office Suits |

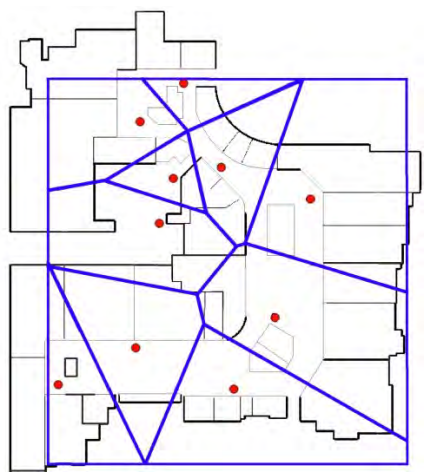
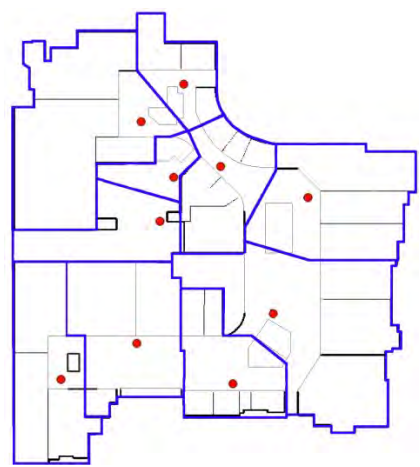

(a) Theoretical Voronoi cells     (b) Rectified Voronoi

Figure 1. An example of APs and the corresponding Voronoi cells. The black lines show the outlines of the stores. The red dots denote the Wi-Fi APs, and the blue lines show the Voronoi cells.

The dataset consists of three kinds of logs: a Wi-Fi Access-point association Log (AL), a Web Browsing Log (BL) and a Web Query Log (QL), collected between September 2012 and October 2013 (Table 2). Before analysis, all user identifiable information in the logs (e.g., user device's MAC address) was replaced by a hash key in an irreversible way. *Users* is the term we use further in this paper to refer to devices appearing in AL, a subset of such users are *browsers* who appear in the BL, and *searchers* are those users who appear in the QL.

Table 2. Aggregate statistics of the AL, BL and QL

| Wi-Fi Access point Log (AL) | |
| --- | --- |
| Number of users: | 120,548 |
| Number of AP association: | 907,084 |
| Number of User Visits: | 261,369 |
| **Web Browsing Log (BL)** | |
| Number of users browsing: | 70,196 (58.3% of AL users) |
| Number of issued URLs: | 18,088,018 |
| Number of User Visits: | 139,004 |
| **Query Log (QL)** | |
| Number of users searching: | 11,169 (9.3% of AL users) |
| Number of queries: | 119,196 |
| Number of query sessions: | 20,637 |

While Web *browsing* is the overall activity detected by recording the user's interaction with Web URLs, the Web *search* activity is detected by recording the visits to URLs following a keyword-based search through a Web search engine. Note that we use a narrower definition of *search* than that applied in (Hodkinson, Kiel, & McCollKennedy, 2000) and restrict this term only for search-engine based search.

## Characteristics of the datasets

**The Wi-Fi AP Association Log (AL)**. The AL captures information about user physical behavior characterized by the following parameters (1) user device's MAC address uniquely identifying the associated device; (2) the users' IP address; (3) the ID of the Wi-Fi access point (not MAC address) associated with the user's mobile device at a given point in time, used as a proxy for the user's location; (4) the time-stamp of users' association/disassociation with the access point; (5) the duration of users' association with the access point; (6) additional parameters (e.g., Received Signal Strength Indication), which are not used further in the scope of this paper.

**The Web Browsing Log (BL)**. The BL includes the users' Web information behavior, characterized by: (1) the time-stamp of the Web request; (2) the users' IP address; (3) the Web page requested, as defined by the URL. This contains all out-going URL requests from the device, including app traffic.

Following (Kumar & Tomkins, 2010; Song, Ma, Wang, & Wang, 2013; Church et al., 2007), we define a browsing session as *a series of URL requests by a single user delimited by 30 minutes of inactivity on the Web*. The duration of a session is defined as the time period between the first and the last URL in the session. We assume that the time within a session is spent on the Web and the time between sessions is not. For user visits accessing only a single URL (around 2.6% of overall user visits in BL) the duration is not defined and they are not further considered.

We enriched the BL by adding a fourth attribute, identifying the location of the user at the time of the request by the ID of the AP. This was done by joining the BL with AL records through a composite key consisting of time-stamp and IP address, recorded in both logs. Note that the first appearance of a users' device in the AL, as well as any consecutive appearance after disconnection always precede the appearance in the BL. It is also possible for the user to only connect to the Wi-Fi network and not access Web pages, thereby only appearing in the AL.

**The Query Log (QL)**. The QL was extracted from the BL by identifying URL requests associated with search engines, including Google (110148, 92.4% of QL), Yahoo (6915, 5.8% of QL), Bing (954, 0.8% of QL), Baidu (1086, 0.9% of QL), AOL (43, 0.04% of QL) and ASK (50, 0.04% of QL). The QL was processed as follows: (1) search queries were treated as case insensitive; (2) a query term was defined as any unbroken string of characters in a query delimited by white-space; (3) the concept of sessions was applied consistently with the processing of the BL.

*Limitations of the datasets*

The logs contain tracking data of mobile devices associated with the Wi-Fi network, by storing the device's (anonymized) MAC address. We assume a MAC address remains representative of a single user across the study. Our AL data capture the timestamp of each device association with a given AP, but movement inside the region served by an AP is not captured. We define a user visit as the combination of all AL records from the same device on a single day. Thus, we assume that multiple users do not share devices at least within the same day.

Only those devices associated with the free Wi-Fi network provided by the mall are logged. This means that a user with a registered device may not be present in a log if they did not associate with the Wi-Fi network studied, but maybe another free Wi-Fi in the mall (e.g., a fast food chain's network) or their own cellular data. In addition, we do not have access to any demographic information about the users and the reasons they visit the mall (e.g. shoppers or mall employees).

As current smartphones typically disassociate from Wi-Fi within a few seconds after the sleep mode turns on, disassociations are frequent and the tracking of users in the mall is not continuous. If a user visits the mall and their device does not associate with the Wi-Fi network, they are not logged. However, many apps send out URL pings frequently, thus keeping smartphones connected.

In the QL, only queries that have been executed over an `http` connection could be analyzed. This is in particular important for Google, which rolled out default encryption of its queries from late 2011[1]. As our dataset originates *after* the roll out, the amount of detected queries may vary, and is likely to decrease with time in our dataset.

# Definitions and Terminology

We define the terminology as used in this paper.

*Physical behavior*

We study the spatio-temporal characteristics of the physical behavior of mall visitors. Their physical behavior largely equates to way-finding activity and may have goal oriented (roaming) and directed search aspects (Wiener, Büchner, & Hölscher, 2009). We restrict our focus on the manifested locomotion of the visitor but in future work hope to be able to detect the nature of the locomotion captured in the data. We denote $A = \{a_1, \ldots, a_m\}$ as the set of all available Wi-Fi APs, where $m$ is the number of APs.

**Definition 1.** *The user's physical behavior during a single visit $v$ is captured by their trajectory, which is expressed as a vector $\mathbf{P}_v$ of the durations $p_{vk}$ that the user spent associated with an AP $a_k$ during the* visit: $\mathbf{P}_v = [p_{v1}, \ldots, p_{vk}, \ldots, p_{vm}]$. *If a user was associated with an AP multiple times in a visit, the total duration of time spent at this AP is stored, while for unvisited APs, the duration is zero.*

### Web behavior

Bates (2010) defined information behavior as the ways in which people interact with information, particularly in terms of how to seek and utilize information. We define the indoor users' Web behavior from two aspects, visits and indoor locations (captured through AP association). We restrict our focus on the subset of information needs that are satisfied through Web interaction, and are unable to consider other social or physical information sources.

We denote $C_w = \{c_w^1, \ldots, c_w^n\}$ as the set of all Web page categories, where $n$ is the number of categories. In this paper, we applied the Web categories defined by the Webroot Content Classification Service (WCCS), BrightCloud (http://bcws.brightcloud.com)[2]. We define two kinds of user Web behavior. First, the behavior during a visit $v$, denoted as $\mathbf{W}_v$:

**Definition 2.** $\mathbf{W}_v$ *is defined as a vector of the number $W_{vk}$ of URLs that are issued during $v$ and belong to $c_w^k \in C_w$:* $\mathbf{W}_v = [W_{v1}, \ldots, W_{vk}, \ldots, W_{vn}]$.

Second, the behavior at a given AP $a_i$ (the overall average Web behavior at an AP), denoted as $\mathbf{B}_i$:

**Definition 3.** $\mathbf{B}_i$ *is defined as a vector of the average number $b_{ik}$ of URLs that are issued through AP $a_i$ and belong to $c_w^k \in C_w$:* $\mathbf{B}_i = [b_{i1}, \ldots, b_{ik}, \ldots, b_{in}]$.

### Physical contexts

We define physical contexts in terms of shop categories (a list of categories for each shop was provided to us by the mall owners), and denote $C_s = \{c_s^1, \ldots, c_s^h\}$ as the set of all shop categories, where $h$ is the number of categories. Then, we denote the spatial indoor context for each AP as $\mathbf{E}_i$:

**Definition 4.** $\mathbf{E}_i$ *is defined as a vector of the number $e_{ik}$ of shops that are located in the Voronoi regions of AP $a_i$ and belong to $c_s^k \in C_s$, giving* $\mathbf{E}_i = [e_{i1}, \ldots, e_{ik}, \ldots, e_{ih}]$.

Vector $\mathbf{E}_i$ is computed for each AP through a spatial overlay operation between the Voronoi region and the outline of shop footprints from the mall floor layout.

### Social contexts

When users are visiting the mall, they may be accompanied by others. To investigate how any social relationship relates to information behavior, we define *social context* by focusing on users with a highly correlated physical behaviors. We define a pair of users as *accompanying* if they: 1) both appear in the AL associated with the same AP ±1 min; 2) there is a >90% overlap in the time recorded in the AL over one visit; 3) at least three different APs are recorded in the AL for both users; 4) the average distance between the users during their visits should be no more than one AP, which means they access the Wi-Fi network via, at most, adjacent AP.

**Definition 5.** *The topological distance between two user visits $v_i$ and $v_j$ is defined as the average step-distance between access points in the Wi-Fi signal topology, with which they are associated during their overlapped visiting time:*

$$d(v_i, v_j) = \frac{\sum_t d(a_{ik}^t, a_{jl}^t)}{\sigma},$$

where $\sigma$ is the overlapping time between $v_i$ and $v_j$ in seconds, $d(a_{ik}^t, a_{jl}^t)$ denotes the topology distance at time $t$, when these two users are visiting AP $a_k$ to $a_l$, respectively.

We focus on users recorded in the AL during opening hours of the mall. We are measuring a topological (step) distance in a graph representation of adjacencies of the service areas of APs as metric distance between the actual positions of users cannot be calculated from the log.

# Basic Behaviors of Indoor Visitors to Retail Environments

Here we describe an overview of the indoor physical and Web behavior of visitors.

## Basic indoor physical and social behavior

**Temporal patterns of users' visits.** We find that the use of the Wi-Fi network maps to the opening hours of the mall. Fig. 2 shows the distribution of the fraction of users' associations/disassociations with the Wi-Fi network (over all users) for each hour in a day for the entire data-collection period. Specially, the red line shows the association trend, and the blue line shows the disassociation trend. Starting from 09:00, the fraction of associations with the network for each hour in a day begins to increase quickly, peaking (12.69%) at 14:00, then begins to decrease until the end of the day. Complementary to the association trend, we can also determine when users last accessed the network (a disassociation). We find that there are more users associating than disassociating with the Wi-Fi network before 15:00, with disassociations peaking at 17:00, around the time when the mall is about to close. The difference between the associations and disassociations enables us to estimate the number of Wi-Fi users in the mall. Note that the differences relate to the percentages, not absolute values.

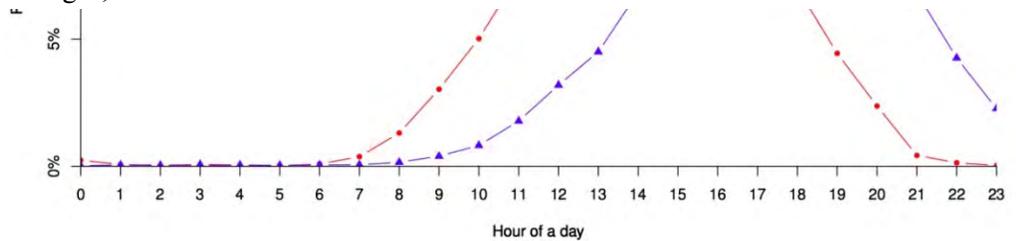

Figure 2. User Wi-Fi association/disassociation distribution over 24 hours.

Moreover, Fig. 3 shows the distribution of user visits over the day of a week. *Thursday* is the most popular day of the week for visiting the mall (17.09%), forming a group of shopping days with *Friday* (15.24%), *Wednesday* (14.50%), and *Saturday* (14.45%). *Monday* (13.57%), *Tuesday* (13.11%) and *Sunday* (12.04%) represent days with lower activity. Thursdays are the typical shopping day in Australia, given the extended opening hours. We note that this mall is a city center mall, and the values may be different at suburban malls.

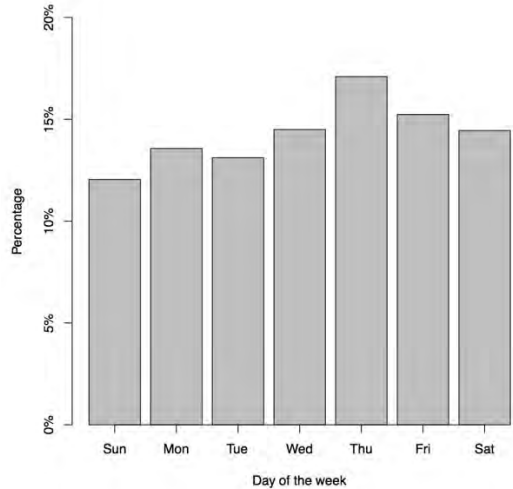

**Figure 3. The user visits over the day of a week.**

**Length of visits to the mall.** We define the detectable visit duration as the period between the first Wi-Fi association and the last Wi-Fi disassociation for each user (device) on any day. We do not make any assumptions about the entire duration of the visitor's stay in the mall beyond the duration captured by Wi-Fi use. Naturally, people may be turning their device on or off during a longer visit, or even not turn their device on when they first enter the shop, or disconnect a significant amount of time before they physically leave – as discussed before. Thus, we only apply the Wi-Fi association time as a *proxy* to the visit duration for a user on a given day.

Overall, indoor users stay in the mall for 2.77 hours on average, with the minimum duration of 0.08 hour and the maximum duration of 21.67 hours. Fig. 4 shows the distribution of the visit durations in our dataset. Around 66% of user visits lasted between three and four hours; 17% lasted less than one hour and around 10% lasted between one and three hours. There is only 7% of user visits lasted more than four hours.

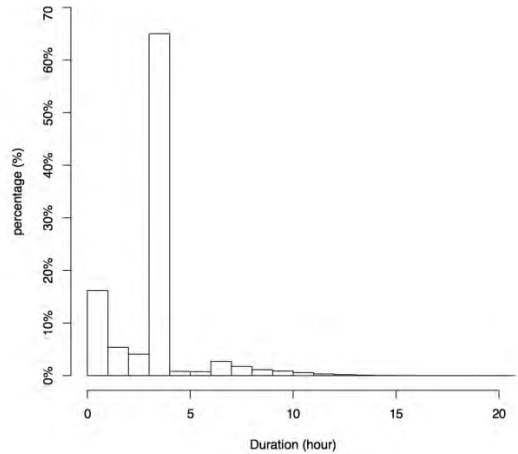

Figure 4. Detectable visit duration (in hours)

**Frequencies of repeat visits.** People have habits that lead to highly repetitive and ultimately predictable patterns (De Domenico, Lima, & Musolesi, 2013; Schulz, Bothe, & Körner, 2012). Here, we explore whether such regularities are present also in the repetitive patterns of returns to a retail environment, hinting at the satisfaction of repetitive needs.

About 67% of users only used the Wi-Fi network once in the monitored period. Of the rest, Fig. 5 shows the distribution of the kinds of user visits classified in function of the difference in days between two consecutive visits of the same user, and we observe that the distribution of the return visits does not follow an uniform decreasing pattern, but a strong impact of a seven-day periodicity is captured in the data (note the local maxima clustered around the multiple of 7-day differences). We cut off the displayed periodicity at 100 days, as the return period is increasing influenced by the data collection period. A consequent question is then whether this visiting repetition pattern correlates with users' physical and Web behavior inside the mall. We analyze this in the following sections.

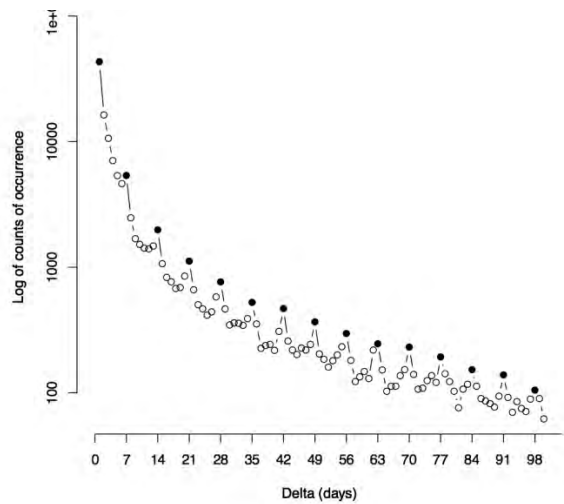

Figure 5. Counts of consecutive visits of all visitors binned by the Δ in days.

**Length of indoor trajectories.** The trajectory length of users is on average 3.47 (expressed in number of APs associated with a visitor per day), with a range of 1-64 (mode=1, median=2). We observe that around 28% of user visits accessed Wi-Fi at a single AP, and the majority (over 93%) of user visits associated with less than 10 APs overall.

**Places of first association.** The place of first association (identified by the AP ID) is not necessarily the same as the point of entry into the mall. We hypothesize that visitors associate either when satisfying direct information needs (either mall related, e.g. price comparison; or generic, e.g. mail checking), or when filling time – eating, resting or waiting for acquaintances. The type of places where the users associate the Wi-Fi network for the first time is therefore informative.

We classify the proximal areas of APs into three contexts:
- *Food-court* context (11 APs, around 16%) denotes the food-court of the shopping mall. The motivation for this direction is that, 1) this is one of the most distinctive directions we observed in the collected data, and 2) we hypothesize that accessing the Web while eating at food court is common for people in indoor retail spaces.
- *Retail* context (46 APs, around 69%) denotes areas covered by APs serving retail areas in the mall (including entertainment types of services).
- *Navigational* context (10 APs, around 15%) covers non-retail areas, e.g. near lifts and escalators, and toilets.

For APs serving spaces with both a retail and navigational context, the AP is classified based on which context covers over 50% of the Wi-Fi signal coverage.

Table 3 shows the distribution of first associations per context, with the average number of Wi-Fi associations per AP in each context. It is observed that only 6% of user association starts in a navigational context with 15% of APs, around 31% starts in the Food-court with 16% of APs, and the majority (63%) starts from a retail context with 69% of APs. However, the number of first Wi-Fi associations per AP is higher in the food-court.

**Table 3. Context of first association**

| Context | % of starting association |
|---|---|
| Food-court | 31% (2.84% per AP) |
| Retail | 63% (1.37% per AP) |
| Navigational | 6% (0.60% per AP) |
| Total | 100% (1.49% per AP) |

Table 4 shows the distribution of visiting time per context. A similar trend to first associations is observed: 7% of users' visiting time are spent at the 15% of APs in navigational areas, 23% of their time are spent at another 16% of APs in food court, and the rest 70% of their time is spent in retail context covered by 69% of APs. Again, the largest average duration per access point is measured in the food-court. In addition, from the average of visiting time per user visit, we observe that indoor users tend to spend more time in *retail* context than other physical contexts in retail environment.

Table 4. Context in relation to visiting time, as a proportion of all association time spent at a given category of AP (and per AP within category), as well as average time.

| Context | % of assoc. time | Avg. time per visit [h] |
|---|---|---|
| Food-court | 23% (2.06% per AP) | 1.39 |
| Retail | 70% (1.52% per AP) | 2.29 |
| Navigational | 7% (0.68% per AP) | 1.00 |
| Total | 100% (1.49% per AP) | 2.77 |

**Basic social behavior.** We now describe the basic aspects of the social behavior identified in the data. We identified 2,705 accompanied user visits, coming from 2,358 individual users, with the size of groups ranging 2-14, and its distribution is shown in Fig. 6. Specifically, the majority (78%) of such visits are composed of 2 users, 15% are composed of 3 users, 4% are composed of 4 users and only 3% are composed of 5 or more users. In the following discussion section, we analyze how accompanying users actively access content on the Web. Identified users who only appear in the AL and not the BL are excluded from this analysis, leaving 2,174 accompanied user visits from 1,886 individual users.

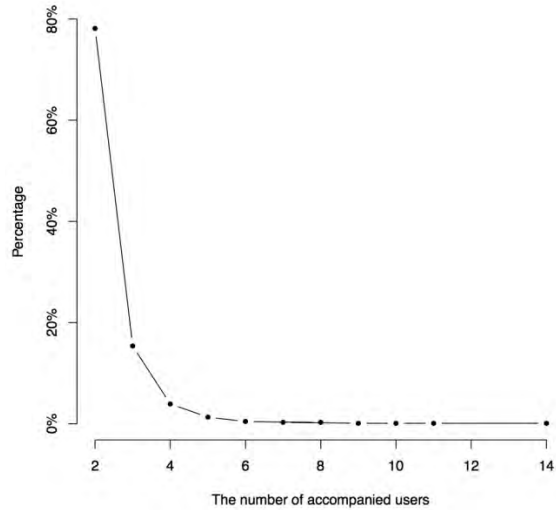

Figure 6. Distribution of the size of accompanied user visits

## Basic indoor Web behavior

**Length of Web access.** Fig. 7 shows the distribution of Web access durations across user visits as captured in BL, via the concept of browsing sessions as defined in this study. The average Web access duration is around 40 minutes, 82% of users accessed the Web for less than an hour. Note the contrast with the distribution of physical visiting time (AL), which showed that 66% of users stayed in the mall between 3-4 hours. Specifically, we observe that the access duration of around 62% of browsers is less than 0.5 hour, 20% of browsers access the Web for a duration of 0.5 to 1 hour, 8% of browsers access for a duration of 1 to 1.5 hours, 4% of browsers access for a duration of 1.5 to 2 hours, and 6% spend more than two hours using the Web in the retail environment.

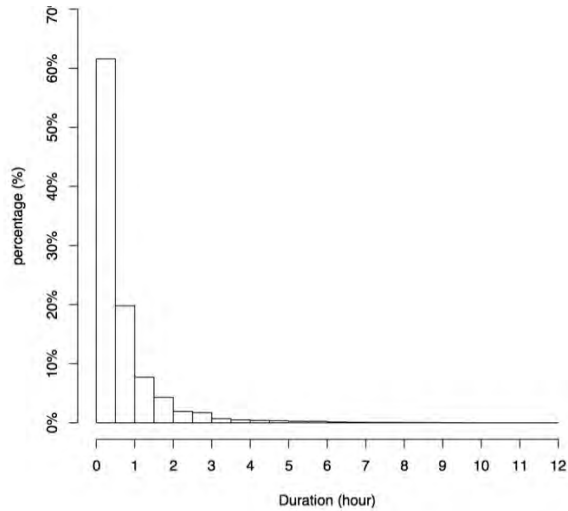

Figure 7. Time spent online

Fig. 8a shows the average BL Web duration, the average AL duration (in range of 0 to 4 hours in hourly bins) and the ratio between these visit durations. While the physical durations of visits (AL) in the mall vary widely, BL durations are much more constraint in extent. On average, a user accessed the Web for less than 1 hour during a single visit, resulting in a decreasing ratio between the BL duration and the AL duration in a visit. This indicates that indoor users are likely to spend a relatively constant amount of time browsing the Web (less than 1 hour), although this period may be fragmented into a number of Web browsing sessions (the average number of sessions per visit is 1.32).

The graph in Fig. 8b shows the distribution of different groups of users (by visit duration revealed through AL) per time spent online. We show users with visits of online duration of 0-0.5 hour, 0.5-1 hours, 1-1.5 hours, and 1.5-2 hours in function of their total time spent browsing. We observe that a large proportion of users in each bar are those who visited the mall for 3-4 hours (as detected from AL). This is expected, as users visiting 3-4 hours in AL take a large proportion of overall users. This group of users is then dominant in each of the browser categories.

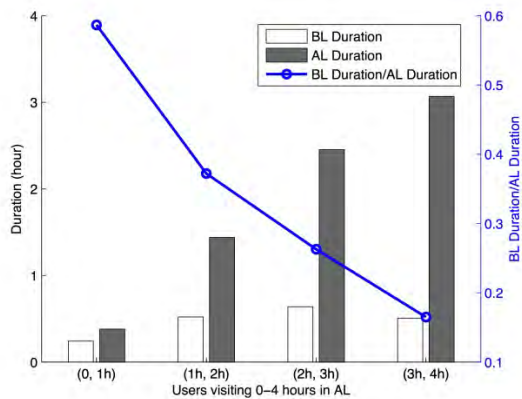

(a) The average BL/AL duration of users visiting 0-4 hours in AL

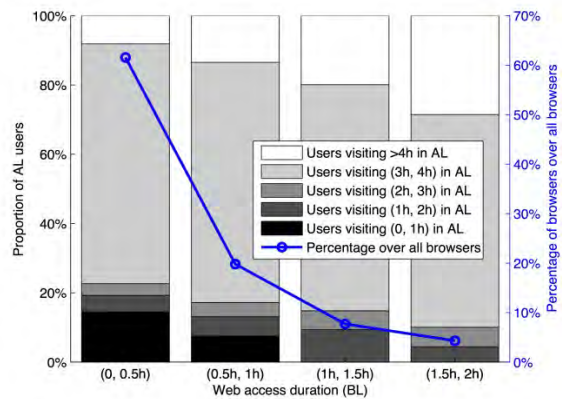

(b) Stacked plot of the composition of

Figure 8. Time spent online (from BL) by user group (identified from AL)

**Content accessed on the Web.** We now analyze what users browse and search for in the mall. A categorization of Website content (captured by URLs, see Table 5, which was briefly discussed in (Ren, Tomko, Ong, & Sanderson, 2014)) was performed using BrightCloud (http://bcws.brightcloud.com).

Table 5. Top 10 categories of browsing and searching

| Browsing | Searching (query-click) |
|---|---|
| Social Networking (20%) | Travel (12%) |
| Content Delivery Networks (13%) | Entertainment & Arts (9%) |
| Computer & Internet Info (12%) | Society (8%) |
| Search Engines (11%) | News & Media (8%) |
| Business & Economy (10%) | Shopping (8%) |
| Personal Storage (5%) | Reference & Research (7%) |
| Web based email (3%) | Social Networking (6%) |
| Web Advertisement (3%) | Business & Economy (6%) |
| News & Media (3%) | Personal Sites & Blogs (4%) |
| Internet Portals (2%) | Computer & Internet Info (4%) |

We observe that *Social Networking* is the most popular browsing category (20%), consistent with overall mobile Web usage (Church & Oliver, 2011). *Content Delivery Networks* (aiming to improve the performance of Web services, e.g., akamaihd.net) and *Computer and Internet info* (e.g., amazonaws.com) take roughly the same proportion, around 13%. *Search Engines* are the fourth most popular category at 11%, and followed by *Business and Economy* with 10.6%. Beyond the dominant category, these values are strikingly different to previous studies focusing on general mobile browsing (Church & Smyth, 2009). We discuss this difference in details in the following discussion section.

We further investigate what users search for in the mall by analyzing Google search results that were followed by the users (query-click). Specifically, browsing categories is derived from all URLs in BL while Searching categories are from the click through from Google's Search Engine Results Page (SERP) in BL. Table 5 captures the Browsing and Searching (query-click) behavior from BL. *Travel* is the most popular category for Searching but only accounts for 1.4% in Browsing; *Social Networking* takes 20% in Browsing but only 6% in Searching.

Fig. 9 shows the cumulative percentage of top Web categories for indoor browsing and searching, indicating the heavy-tailed distribution pattern in Web content access. Specifically, around 80% of indoor Web browsing and searching URLs come from the top 20% of Web categories – showing a typical long-tail distribution characteristic. We discuss the implications of these findings in the following discussion section.

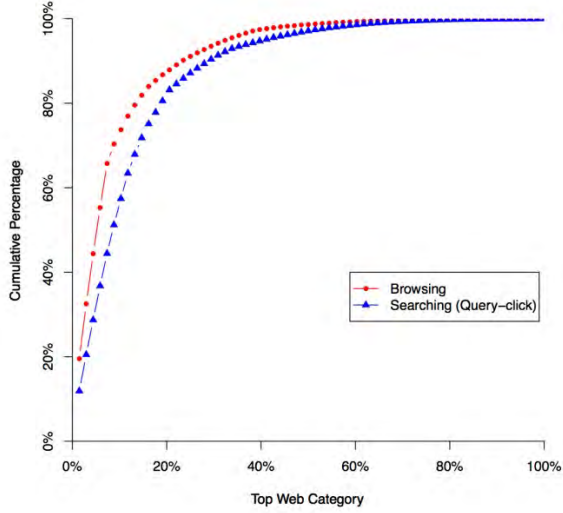

Figure 9. Browsing vs. Searching in terms of cumulative percentage of top Web categories

# Analyzing Indoor Behavior

*Commonly accessed web content*

We use the concept of entropy to quantify the commonality of a Website category in the Web behavior of users by measuring the *access entropy* across users.

For a URL category $c_w$, access entropy $H(c_w)$ is defined as:

$$H(c_w) = - \sum_{v \in S(c_w)} p(v|c_w) \log p(v|c_w), \quad (1)$$

where $S(c_w)$ is the set of visits when users accessed URLs in category $c_w$, $p(v|c_w)$ is the percentage of accesses to $c_w$ during a visit $v$ out of all visits. A high access entropy $H(c_w)$ means that $c_w$ is a common category among all users; a low entropy means a category is accessed by a subset of users. Fig. 10 shows the distribution of $H(c_w)$. *Computer and Internet Info*, *Social Networking* and *Search Engines* are common URL categories with entropies of 10.75, 10.72 and 10.50, respectively. We observe that there are some categories of Websites that are more commonly visited than others, and given the $x$-axis ($H(c_w)$) of Fig. 10 is on a log (bits of entropy) scale, we conclude that there is a small number of categories that dominate what user access on the Web.

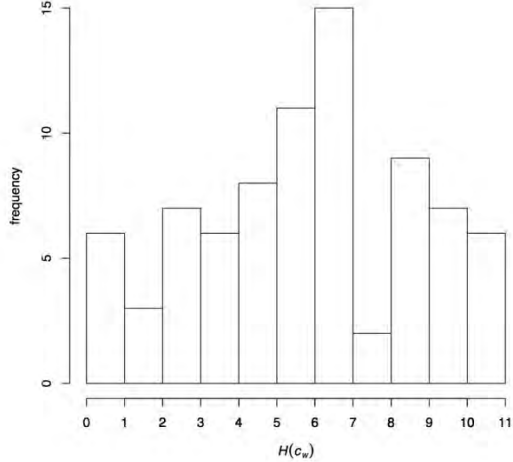

Figure 10. Binned distribution of access entropy $H(c_w)$

Some categories are commonly issued by a large proportion of users during many user visits, but they are not high in absolute numbers in the overall URL traffic. For example, the category *Shareware & Freeware* covers URL requests to Web pages containing screensavers, icons, wallpapers, utilities and ringtones. These are commonly accessed (high $H(c_w)$). However, the absolute number of requests to such Web URLs is low and this category does not even rank in the top 10 popular categories in Table 5.

*Examining repetitive patterns*

We investigate the stability of users' indoor behavior during consecutive visits. To measure the strength of the correlation of the physical behaviour $\mathbf{P}_{v_i}$ of a user during a current visit $v_i$ with the behavior $\mathbf{P}_{v_j}$ during a consecutive visit $v_j$ (further called *repeat model*), we compute the Pearson Correlation Coefficient (PCC) over consecutive visits as:

$$PCC_{phy} = r\left(\mathbf{P}_{v_i}, \mathbf{P}_{v_j}\right) = \frac{\sum_{a_k \in A}(p_{ik} - \bar{p}_i)(p_{jk} - \bar{p}_j)}{\sqrt{\sum_{a_k \in A}(p_{ik} - \bar{p}_i)^2 \sum_{a_k \in A}(p_{jk} - \bar{p}_j)^2}}, \qquad (2)$$

where $\bar{p}_i$ and $\bar{p}_j$ are the average duration a user spends at each access point for visit $v_i$ and $v_j$, respectively. A high positive PCC value indicates a strong correlation in physical behavior during consecutive visits. We apply two baselines to compare with the repeat model: a *random*-paired baseline and an *average* baseline: the former replaces each repeated $\mathbf{P}_{v_j}$ with a randomly selected visit, and the latter replaces with the average physical behavior $\overline{\mathbf{P}}_v$ over all user visits[3]:

$$\overline{\mathbf{P}}_v = \frac{\sum \mathbf{P}_v}{|V|},$$

where $V$ is the set of user visits and $|V|$ is the number of the user visits.

Table 6 shows the PCC values for *repeat*, *random*, and *average* models. It can be seen that *repeat* achieves the largest PCC value, which is over two times larger than that of *average* and over twenty times larger than that of *random*. We have analysed the variance between the means of the *repeat*, *random*, and *average* models through ANOVA and conclude that the differences are statistically significant with a *p*-value of < 0.0001. This indicates that users' physical indoor behavior is repetitive and does not change substantially between two consecutive visits. It demonstrates that visitors return to the same parts of the mall and spend similar amounts of time in them.

Table 6. PCC values of trajectories

|  | Repetition | Random Pair | Average |
|---|---|---|---|
| All | 0.2534 (±0.3922) | 0.0123 (±0.1506) | 0.1108 (±0.1585) |

Similarly, we apply PCC to measure the correlation in Web behavior between two consecutive visits $v_i$ and $v_j$,:

$$PCC_{web} = r\left(\mathbf{W}_{v_i}, \mathbf{W}_{v_j}\right).$$

Again, we define another two baselines: the *random*-paired baseline, which replaces $\mathbf{W}_{v_j}$ with a randomly selected visit, and the *average* baseline, which replaces $\mathbf{W}_{v_j}$ with the average Web behavior $\overline{\mathbf{W}}_v$:

$$\overline{\mathbf{W}}_v = \frac{\sum \mathbf{W}_v}{|V|},$$

where $V$ is the set of user visits and $|V|$ is the number of the user visits.

The first row of Table 7 shows the PCC results when all Web categories are considered, including those with a high access entropy $H(C_w)$. We observe that consecutive visits achieve the highest score $r = 0.5902$, which means they are highly similar; *average* follows with $r = 0.5068$ while random only reaches $r = 0.2647$. To show the positive correlation between Web accesses in consecutive visits more clearly, we gradually remove common Web categories by setting a threshold for $H(C_w)$[4], and then re-calculate the above experiments (Table 7). The gap between *repetition vs. random* and *repetition vs. average* increases when the common Web categories are gradually removed. A two-tailed, paired $t$-test was applied to evaluate whether the differences are statistically significant (results in Table 8). It indicates that the PCC values for *repeated* visits are statistically larger than both that for *random* and *average*.

Table 7. PCC values of browsing log (over Brightcloud category) for consecutive visits, random paired visits and between (each visit, average visit profile)

| $H(c_w)$ | Repetition | Random Pair | Average |
|---|---|---|---|
| $H(c_w) \leq max(H(c_w))$ | **0.5902** | 0.2647 | 0.5068 |
| $H(c_w) \leq 10$ | **0.4581** | 0.0922 | 0.3010 |
| $H(c_w) \leq 9$ | **0.4311** | 0.0694 | 0.2632 |
| $H(c_w) \leq 8$ | **0.5261** | 0.0287 | 0.1875 |
| $H(c_w) \leq 7$ | **0.4940** | 0.0236 | 0.1505 |
| $H(c_w) \leq 6$ | **0.6526** | 0.0483 | 0.2422 |
| $H(c_w) \leq 5$ | **0.7986** | 0.1096 | 0.2093 |

Table 8. Paired $t$-test results for PCC values of browsing log comparison

| Methods | Paired-$t$ statistics | |
|---|---|---|
|  | $t$ | $p$-value |
| Repetition vs. Random | 8.2 | < 0.0001 |
| Repetition vs. Average | 4.545 | 0.0007 |

We have also examined the browsing differences between different visiting periodicities. We find that as time between revisits increases, there is decay in the likelihood of users repeating what they looked at online compared to last time. The PCC values degrade from around 0.63 for a periodicity of one day to about below 0.55 for a periodicity of 6 days, with a small increase in around 7 days.

*Spatial context & information behavior*

There are differences in the categories of shops served by different Wi-Fi access points (the association being done using the discussed Voronoi regions). We hypothesize that the proximity of different shop categories (the indoor context) will lead to a different Web information behavior of the mall visitors. At the level of Wi-Fi APs, the influence of spatial context on users' Web behavior can be viewed as the correlation between $\mathbf{B}_i$ and $\mathbf{B}_j$ for every two APs. We again apply PCC to test this association:

$$r(\mathbf{B}_i, \mathbf{B}_j) = \frac{\sum_{c_w^k \in C_w}(b_{ik} - \bar{b}_i)(b_{jk} - \bar{b}_j)}{\sum_{c_w^k \in C_w}(b_{ik} - \bar{b}_i)^2 \sum_{c_w^k \in C_w}(b_{jk} - \bar{b}_j)^2}, \quad (3)$$

where $C_w$ is the set of URL categories, $\bar{b}_i$ and $\bar{b}_j$ are the average numbers of issued URLs at $a_i$ and $a_j$, respectively.

**Influence of location on Web behavior.** To test the above hypothesis, we analyze the *average* PCC value $r$ between every pair of access points from Eq. 3, and it is defined as:

$$average = \frac{2}{|\mathfrak{B}|(|\mathfrak{B}|-1)} \sum_{\mathbf{B}_i} \sum_{\mathbf{B}_j, i \neq j} r(\mathbf{B}_i, \mathbf{B}_j),$$

where $\mathfrak{B}$ denotes the set of user Web behavior, and $|\mathfrak{B}|$ denotes the size of $\mathfrak{B}$. Results are shown in Fig. 11 and the rightmost column of Table 9. The average of $r$ reflects the general similarity of Web activity throughout the space, with a small $r$ indicating that different locations in the mall lead to different user Web behavior. Using all URL categories, the average value of $r$ is 0.9619, indicating that there is little difference between the Web behavior at different APs. However, this correlation is caused by the large proportion of common Web requests pointing to a small subset of URLs of well-defined categories. The top 5 common URL categories, which are identified by access entropy $H(c_w)$, take over 57.8% of the overall URL records and thus dominate the dataset. This significantly skewed Web behavior introduces a bias in the $r$ value.

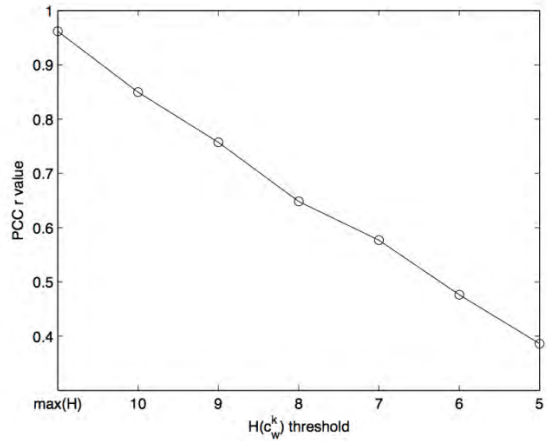

**Figure 11. PCC $r$ value without common $c_w$**

Table 9. Correlation of user Web behavior in groups of access points with similar spatial context

| | $H(c_w)$ | PCC $r$ value based on $\mathcal{B}$ | | | | | | average |
|---|---|---|---|---|---|---|---|---|
| | | $k$-means | | places | | random | | |
| | | within | between | within | between | within | between | |
| Groups of Access Point based on $\mathcal{E}$ | $H(c_w) \leq \max(H(c_w))$ | **0.9659** | 0.9623 | 0.9617 | 0.9613 | 0.9609 | 0.9617 | 0.9619 |
| | $H(c_w) \leq 10$ | **0.8601** | 0.8509 | 0.8401 | 0.8302 | 0.8493 | 0.8501 | 0.8498 |
| | $H(c_w) \leq 9$ | **0.7721** | 0.7599 | 0.7540 | 0.7287 | 0.7564 | 0.7573 | 0.7573 |
| | $H(c_w) \leq 8$ | **0.6817** | 0.6572 | 0.6804 | 0.6556 | 0.6493 | 0.6473 | 0.6483 |
| | $H(c_w) \leq 7$ | **0.6410** | 0.5966 | 0.5950 | 0.5645 | 0.5767 | 0.5750 | 0.5770 |
| | $H(c_w) \leq 6$ | **0.5045** | 0.4778 | 0.5001 | 0.4842 | 0.4755 | 0.4751 | 0.4763 |
| | $H(c_w) \leq 5$ | **0.4107** | 0.3942 | 0.4004 | 0.3837 | 0.3821 | 0.3848 | 0.3863 |

We therefore conduct another experiment to isolate the influence of the frequent Websites on the detection of Web activity. The assumption is that those common Websites represent a baseline user behaviour that is not specific to, or determined by the indoor context. We remove top common URL categories identified by Eq. 1 based on $p(v|c_w)$, which means the identification of URL commonality is based on user visits $v$ and is unrelated with physical context. As such, the identification of URL commonality is independent from the subsequent calculation of $r$.

To show the influence of indoor location on user Web behavior, we calculate the $r$ value by progressively eliminating common URL categories. We select $c_w$ based on its access entropy $H(c_w)$, with a threshold and vary the threshold from $max(H(c_w))$ to 5 with a unit step. The $r$ value is calculated by applying Eq. 3 based on $\mathbf{B}_i$.

When common URLs are removed from the calculation of $r$, differences in Web behavior at different access points appear. The more common URL categories we remove, the more substantial a difference we observe indicating that there is an influence from the local context of access points on user Web behavior.

**Influence of indoor context.** To further investigate regularities in the influence of indoor context, we apply a clustering algorithm to group similar access points based on shop categories. From *definition* 4, an AP $a_i$ is represented by a vector $\mathbf{E}_i$ of shop categories. If the users' information behavior is influenced by their indoor context, the users' Web behavior *within* a cluster should be *similar* and the users' Web behavior *between* clusters should be *different*. We apply the $k$-means clustering algorithm to cluster $\mathcal{E}$ by treating each $\mathbf{E}_i \in \mathcal{E}$ as an instance[5]. We set $k = 6$ because it achieves a relatively low value of the Davies-Bouldin index (Davies & Bouldin, 1979).

To test the association, we apply PCC again to measure the similarity between the Web behavior at two access points. The *intra-cluster* similarity (*within*) and the *inter-cluster* similarity (*between*) are defined as follows:

$$within = \frac{1}{k} \sum_{x=1}^{k} \left( \frac{2}{|t_x|(|t_x| - 1)} \sum_{\mathbf{B}_i \in t_x} \sum_{\mathbf{B}_j \in t_x, i \neq j} r(\mathbf{B}_i, \mathbf{B}_j) \right), \quad (5)$$

$$between = \frac{1}{k}\sum_{x=1}^{k}\left(\frac{1}{|t_x|(|\mathcal{B}|-|t_x|)}\sum_{\mathbf{B}_i \in t_x}\sum_{\mathbf{B}_j \notin t_x} r(\mathbf{B}_i, \mathbf{B}_j)\right), \quad (6)$$

where $k$ is the number of clusters, $t_x$ denotes the $x$-th cluster, and $|t_x|$ denotes the size of $t_x$. We emphasize that the groups of access points are clustered based on their physical context information $\mathcal{E}$, but the $r$ value is defined based on user's Web behavior $\mathcal{B}$. Hence, the user's information behavior is isolated from the clustering process.

We vary $H(c_w)$ from $max(H(c_w))$ to 5 with a unit step. We apply a *random* clustering method as a baseline to show the influence of indoor context. The *average* $r$ for all $\mathbf{B}_i$ pairs is also applied as another baseline. In addition, we examine the influence of the coarser indoor contexts (food-court, retail and navigational) on users' Web behavior. Specifically, we treat this as *places*-based clustering results and calculate the corresponding *within* and *between*.

Table 9 shows the results of the experiment and Table 10 the results of the analysis, where a two-tailed, paired $t$-test is applied to evaluate whether the observed influence is significant or not. We observe: (1) the *within* of $k$-means is significantly larger than the *between* of $k$-means. (2) the *within* of $k$-means is significantly larger than the *within* of *random* and *places*-based methods. (3) the *within* of $k$-means is significantly larger than the *average*. (4) the *within* of *places* is significantly larger than its *between* value, which indicates that the contextual influence is detectable even when the physical contexts are defined at a coarse-grained level. (5) the *within* of *random* is not significantly different from its *between* value. (6) the *within* of *random* is not significantly different from the *average*. As shown in the first row of Table 9, even when no common URL categories are removed, the *within* value of $k$-means 0.9659 is larger than the corresponding *between* value 0.9623, and is also larger than that of *random* 0.9609, *places* 0.9617 and the *average* 0.9619.

Table 10. Paired $t$-test results

| Methods | Paired-t statistics | |
|---|---|---|
| | t | p-value |
| $within$($k$-means) VS $between$($k$-means) | 3.7962 | 0.0090 |
| $within$($k$-means) VS $within$(random) | 3.5871 | 0.0115 |
| $within$($k$-means) VS $within$(places) | 2.5497 | 0.0435 |
| $within$($k$-means) VS $average$ | 3.4126 | 0.0143 |
| $within$(places) VS $between$(places) | 4.5326 | 0.0040 |
| $within$(random) VS $between$(random) | 0.2526 | 0.8090 |
| $within$(random) VS $average$ | 1.6007 | 0.1606 |

The results show that the observed influence is statistically significant (see paired-$t$ statistics in Table 10). This indicates that there is an influence from indoor spatial context on users' Web behavior.

Finally, we examine what Web content indoor users accessed in different context, e.g. retail context, food-court or navigational context, as shown in Table 11. Specifically, around 70% of Web sites about *Swimsuits & Intimate Apparel*, *Fashion and Beauty*, *Alcohol and Tobacco*, *Financial Services*, and *Shopping*, are accessed in the retail context; around 50% of Web pages about *Kids*, *Home and Garden*,

*Real Estate*, *Individual Stock Advice and Tools*, and *Sports*, are requested in the food-court section; *Dating*, *Search Engines*, *Social Networking*, *Web based email*, and *Fashion and Beauty* are popular services accessed by users in the navigational context.

Table 11. Popular Web categories in each context

| Retail | Food-court | Navigational |
|---|---|---|
| Swimsuits & Intimate Apparel | Kids | Dating |
| Fashion and Beauty | Home and Garden | Search Engines |
| Alcohol and Tobacco | Real Estate | Social Networking |
| Financial Services | Individual Stock Advice and Tools | Web based email |
| Shopping | Sports | Fashion and Beauty |

*Social context and Web access*

To investigate what the accompanying users access on the Web, we measure the overlap of the accessed Web content captured through Web domains. For two accompanied user visits, $v_i$ and $v_j$, we define

$$O_{social} = \frac{|D_{v_i} \cap D_{v_j}|}{|D_{v_i} \cup D_{v_j}|}, \quad (7)$$

where $D_{v_i}$ is the set of Web domains that a user visit $v_i$ accessed on the Web.

To show the influence of the social context we compare $O_{social}$ with two baselines:
- $O_{physical}$ : when computing the domain overlap as shown in Eq. 7, replace $v_j$ with another random user visit, which is associated with exactly the same Wi-Fi access points associated by $v_i$. This baseline will distinguish the influence of the accompanying social context from that of the physical context.
- $O_{random}$ : replace $v_j$ with another random user visit when calculating the domain overlap defined in Eq. 7.

Table 12 shows the average values of $O_{social}$, $O_{physical}$ and $O_{random}$ over various groups of accompanying users whose average distance is ≤ 1.

Table 12. Overlap in accessed Web domains amongst members of a group

| $d(v_i, v_j)$ | $O_{social}$ | $O_{physical}$ | $O_{random}$ |
|---|---|---|---|
| $d(v_i, v_j) = 0.0$ | 0.1868 | 0.1119 | 0.1031 |
| $d(v_i, v_j) \leq 0.1$ | 0.1833 | 0.1173 | 0.1057 |
| $d(v_i, v_j) \leq 0.2$ | 0.1780 | 0.1130 | 0.1054 |
| $d(v_i, v_j) \leq 0.3$ | 0.1772 | 0.1139 | 0.1067 |
| $d(v_i, v_j) \leq 0.4$ | 0.1717 | 0.1147 | 0.1060 |
| $d(v_i, v_j) \leq 0.5$ | 0.1670 | 0.1173 | 0.1072 |
| $d(v_i, v_j) \leq 0.6$ | 0.1635 | 0.1137 | 0.1084 |
| $d(v_i, v_j) \leq 0.7$ | 0.1620 | 0.1175 | 0.1061 |
| $d(v_i, v_j) \leq 0.8$ | 0.1628 | 0.1160 | 0.1061 |
| $d(v_i, v_j) \leq 0.9$ | 0.1613 | 0.1162 | 0.1043 |
| $d(v_i, v_j) \leq 1.0$ | 0.1614 | 0.1157 | 0.1049 |

We show that the domain commonality in accompanying users' visits is higher than that modeled by the baselines $O_{physical}$ and $O_{random}$. Moreover, Table 13 shows the paired-$t$ test results among accompanied visits, *physical*-paired visits and *random*-paired visits, and we observe that (1) $O_{social}$ is significantly larger than $O_{physical}$; (2) $O_{social}$ is significantly larger than $O_{random}$; (3) $O_{physical}$ is significantly larger than $O_{random}$, which confirms the influence of physical context. These indicate that the accompanying social context significantly correlates with the Web content consumed during people's visits to indoor retail spaces. In other words, visitors belonging to the same social group access similar content on the Web. Furthermore, we show that this influence is not just an artifact of the joint physical context (proximity to the same shops).

Table 13. Paired $t$-test results for overlap in domains

| Methods | Paired-$t$ statistics | |
|---|---|---|
| | $t$ | $p$-value |
| $O_{social}$ VS $O_{physical}$ | 19.3371 | < 0.0001 |
| $O_{social}$ VS $O_{random}$ | 22.8111 | < 0.0001 |
| $O_{physical}$ VS $O_{random}$ | 13.1395 | < 0.0001 |

Accompanying users are more likely to access the same Web content (domains). Although the extent of this overlap is not large, it is statistically significant. But is this content similar to the overall commonly accessed Web content of indoor users? We first examine the distribution of Web domains in the collected BL (Fig. 12a). The distribution of Web domains is highly skewed and has a long tail. Over 80% of the overall Web accesses go to less than 1% of overall Web domains in the collected data. This is expected following the discussion about basic indoor Web behavior. We investigate what are the commonly accessed Web domains from the accompanying users. Here, we define $D_{social}$ as the union set of domains that are commonly accessed by every accompanied user visits, corresponding to $O_{social}$; and $D_{random}$ as the union set of domains that are commonly accessed by an accompanied user visit and another randomly selected user visit, corresponding to $O_{random}$.

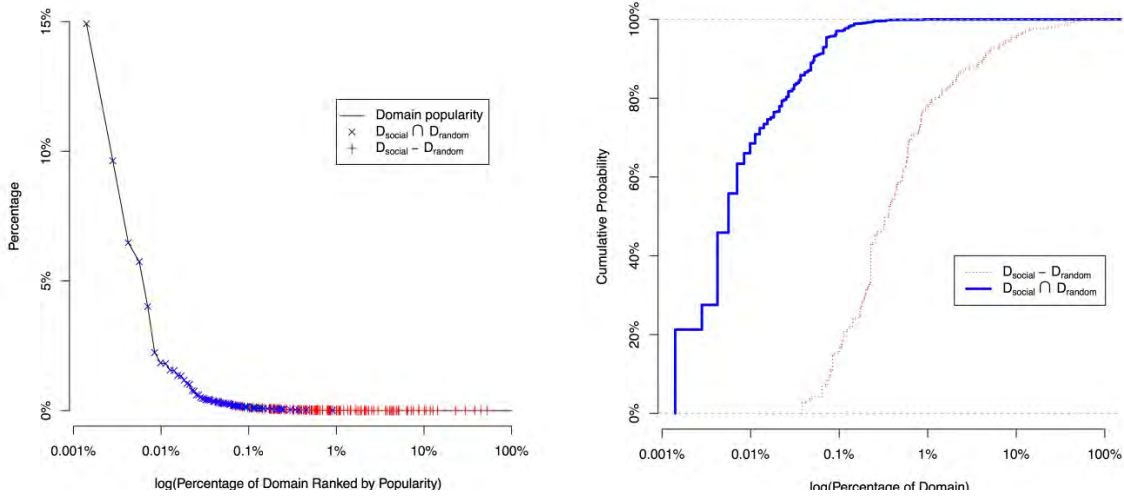

(a) the log plot of domain popularity and the distribution of $D_{social} \cap D_{random}$ and $D_{social} - D_{random}$ over it.

(b) the empirical CDFs of $D_{social} \cap D_{random}$ and $D_{social} - D_{random}$

**Figure 12. The domain popularity and the relationship between $D_{social}$ and $D_{random}$**

Thus, $D_{social} \cap D_{random}$ reflects the domains that are commonly accessed by an indoor user regardless of whether they are accompanied or not, and $D_{social} - D_{random}$ reflects the domains that are shared among accompanying users but not non-accompanying users. Finally, we obtain that $|D_{social}|$ = 208, $|D_{random}|$ = 88, $|D_{social} \cap D_{random}|$ = 70 and $|D_{social} - D_{random}|$ = 138.

The blue-cross points and the red-plus points in Fig. 12a show the distribution of $D_{social} \cap D_{random}$ and $D_{social} - D_{random}$ over the distribution of overall Web domains, respectively. We observe that $D_{social} \cap D_{random}$ are composed of Top popular Web domains, namely Top 1%; $D_{social} - D_{random}$ are composed of unpopular Web domains, which mainly come from the 'tail' of the overall domain distribution. Furthermore, we examine the difference between the distributions of $D_{social} \cap D_{random}$ and $D_{social} - D_{random}$, as shown in Fig. 12b in terms of empirical CDFs. We observe that their CDFs are different, and the Kolmogorov-Smirnov test has been applied to measure whether the differences are significant. The detailed result is (D = 0.8804, *p*-value < 0.0001), which means the differences are statistically significant. This indicates that apart from accessing popular Web domains, the accompanying users tend to access some less popular Web domains which may be specific to their information needs.

A following question is what kind of domains is accessed by social accompanying users. It is observed that these domains belong to both poplar and unpopular Web categories. Specifically, around 65% of accompanying users used the same *Social Networks*, e.g. facebook.com, instagram.com, twitter.com, path.com, renren.com; 25% of them accessed the same *Personal Storage websites*, e.g. icloud.com, dropbox.com, me.com, gogii.com; 20% of them accessed the same *Web based email servers*, e.g., hotmail.com, gmail.com, outlook.com; another 20% of them visited the same *Shopping Websites*(e.g., gumtree.com.au, admob.com, vivant.com.au), *News and Media Websites*(e.g. jyllandsposten.dk, mistermedia.nl, nzherald.co.nz, snstatic.fi, bbc.co.uk, bonzaii.no), *Financial services*(e.g. nordea.dk, westpac.com.au, commbank.com.au, bankofamerica.com, nab.com.au, bango.net, navyfcu.org), and *Games servers*(e.g. king.com); and 15% of them used the same *Internet Portals websites*, e.g. yahoo.com,

live.com, qq.com, sina.cn, msn.com. Please note two accompanying users may access Web domains from more than one category. This shows that accompanying users tend to have similar habits and needs in the mall.

Moreover, we also examine where the accessed domains belong to by checking the corresponding top-level country domain. In particular, it is observed that around 15% of accompanied user visits accessed at least one Website having one same country domain, which are over 26 countries, including au(Australia), us(United States), cn(China), dk(Denmark), nz(New Zealand), se(Sweden), sg(Singapore), nl(Netherlands), co(Colombia), fi(Finland), kr(Republic of Korea), ca(Canada), no(Norway), cc(Cocos Islands), pl(Poland), it(Italy), de(Germany), io(British Indian Ocean Territory), vn(Vietnam), ru(Russia), ie(Ireland), cl(Chile), ch(Switzerland), uk(United Kingdom), am(Armenia and Nagorno-Karabakh Republic). On contrary, only 3% of randomly paired users (corresponding to $D_{random}$) accessed the Websites over only around 5 countries. Please note, for the domain au(Australia), to eliminate the effect of the host location of the investigated mall, we only limit the domain to local services(e.g., dualaustralia.com.au, vodafone.com.au, westpac.com.au) other than localized well-known world-wide services (e.g., google.com.au). This indicates that there is a good probability to see shopping accompanies having the same nationality, assuming people tend to use their native language in daily life.

# Discussion

We now discuss the findings and highlight avenues where deeper insights and research are needed to establish the cause of some of the investigated patterns.

*Temporal patterns of users visits*

The analysis of the length of visits to the indoor environment shows an uneven distribution with the majority of visitors spending 3-4 hours in the mall, while visits shorter than 1 hour are common. The likelihood of a user returning to the mall is higher if the time difference since the last visit is aligned with a weekly pattern or its multiple. These two patterns may point to different *purposes* of the visits to the retail environment and the related nature of the indoor physical behavior. The trip may be related to the satisfaction of repetitive needs, further emphasized by a preference for a specific day of the week for shopping when conducting goal-oriented regular shopping trips (e.g., weekly grocery shopping for locals). Additional rounding on the periodicity capturing individual flexibility in the choice of the day of the week for the shopping trip (e.g., shopping on Fridays or Saturdays) may emphasize this effect. Less regular shoppers may visit on an ad-hoc basis related to an activity satisfying rare needs (e.g. buying a present, cinema visit). These patterns may prove useful for the detection of customer groups. A first venture in this direction is the analysis of the locations that users visit during repeat visits. We show that the closer together visits are in the sequence, the more similar their pattern is likely to be. Combined with a deeper analysis of social shopping contexts and shopper groups, our future research will focus on the predictability and characterization of these groups.

*Spatial patterns of indoor visits*

Recall that the visitors are only monitored if actively interacting with their mobile devices. The short length of indoor trajectories detected might indicate that indoor visitors use Wi-Fi in a relatively static manner, for instance while eating at a food court (phones enter sleep mode when not actively used during walking or shopping). Food courts are also locations of the longest average connection duration per AP and proportionally they are also the most likely place of first association. While about 70% of AP accesses occurs in the normal retail context (naturally as most of the mall is dedicated to this function), a high proportion of Web use occurs in the food-court context. The length of association with APs in the food-court context is also high (on average 1.39 hours). We conclude that the food-court context in shopping malls has a dominant role in the visitors' Web behavior and it is therefore critical to improve the quality of services in this part so as to satisfy users' information needs better.

*Web content use and context dependence*

We found that for the different groups of visitors grouped by length of time spent in the mall, the amount of time spent on the Web does not vary much and is less than an hour for about 70% of visitors. This may indicate that for the majority of indoor visitors, accessing the Web is not the primary activity pursued in the mall and that their information needs satisfied via Web require a relatively constant amount of time, independent of the total amount of time spent in the mall.

We have then examined the content users consume online. Church and Smyth (Church & Smyth, 2009) report only 3.2% for *Email and Social Networking* in general mobile Web access, while our study shows that this category is much more represented (at 23.1%). It is possible that either the indoor context leads to a different information behavior than general mobile Web use, or that the information behavior of mobile users has changed since the publication of the study of Church and Smyth (Church & Smyth, 2009).

There is a pronounced difference between the Web content browsed and searched, with a dominant representation of social networking services *browsed to*, rather than *searched for*. We can hypothesize that with the increased use of smartphones for personal communication, people access emails and similar service much more often and possibly for shorter sessions. Furthermore, this may indicate that the users either know the URLs of social networking services (and other frequently used Web destinations) because they use them routinely and therefore do not need to search for them, or use pre-installed apps to access them. We can hypothesize that the query activity is targeted to satisfy ad-hoc information needs while direct browsing activity may target repetitive needs. This hypothesis is further supported by the fact that social networking as well as email access constitute context independent activities that are part of the regular and frequent Web activity.

We further show that once common Websites are filtered out (the top 5 common URL categories take over 57.8% of the overall URL records, the Web behavior of the visitors reveals strong contextual dependence. Compared to a baseline generated by random and average models, the *repeat model* taking into account the Web content accessed in the previous visit allows for a substantial improvement in content prediction in the consecutive visit. Thus, the content access is correlated in time and space, with different Web content accessed in different parts of the mall, as well as different parts of the mall with the same *context* (mixture of shop categories) inciting users to consume similar Web content. Finally, we have demonstrated how visitors belonging to the same social group have a Web behavior biased to a larger proportion of joint Web content consumed within the mall.

We have thus shown that the visitor's Web and physical behavior is predictable and highly contextualized and can be modeled beyond individual visitors, in visitor groups that can be detected purely based on their spatio-temporal characteristics.

## Conclusions and Future Work

Based on a large data set collected over a one year period through an opt-in public Wi-Fi network of a large urban shopping mall in Australia, we present an analysis of how people use the Web in the context of indoor retail spaces. Specifically, we focus on the following research questions:

- Does the use of Wi-Fi network map the opening hours of the mall?
- Do users tend to visit the retail mall on a certain frequency?
- Are users likely to access the Web while visiting the mall?
- Do users always keep accessing the Web during their visits?
- Do users tend to visit similar mall locations and Web content during their repeated visits to the mall?
- Does users' Web behavior correlate with the indoor spatial context?
- Does users' social context correlate with their Web behavior?

Our findings demonstrate that:

- The use of Wi-Fi network in a retail mall corresponds to the opening hours of the mall;
- There is a weekly periodicity in users' visits to the mall;
- Around 60% of registered Wi-Fi users actively browse the Web and around 10% of them use Wi-Fi for accessing Web search engines, and the content that indoor users search for is different from the content they consume while browsing;
- People are likely to spend a relatively constant amount of time browsing the Web while their visiting duration may vary;
- Users tend to visit similar mall locations and Web content during their repeated visits to the mall;
- The physical spatial context has a small but significant influence on the Web content that indoor users browse on the Web;
- Accompanying users tend to access resources from the same Web domains.

The study established the extent of the predictability of contextualized indoor information behavior, a first step towards visitor modeling. The patterns in suburban shopping malls or in malls in other countries may differ. The study also raised many new research questions: 1) How to improve users' Web experience in the context of indoor retail spaces? 2) What are the specific differences in indoor users' Web behaviors in two kinds of indoor contexts? 3) Can the differences in Web behavior help to identify the spatial context of user preferences, and can this knowledge be utilized further to provide contextual preference-aware recommendations to satisfy user needs? We hope we contributed to a better understanding of people's indoor information behavior in retail environments. As over 80% of shoppers check the price online before purchase (Regalado, 2013), and 27% of smartphone users do research while in store, a better understanding of indoor information behavior can help improve services to shoppers.

## Acknowledgements

This research is supported by a Linkage Project grant of the Australian Research Council (LP120200413).

Footnote:
1. http://googleblog.blogspot.com.au/2011/10/making-search-more-secure.html
2. There are other WCCS, such as DMOZ, but our testing found that its coverage was too narrow for our study. E.g., the highly popular Australian classifieds Website *www.gumtree.com.au* is not categorized in DMOZ but categorized as *shopping* by *BrightCloud*.
3. All random processes in this research are repeated ten times, and averaged.
4. When $H(c_w) \leq 4$, some $W_i$ become empty, which renders the calculation of $PCC_{web}$ undefined. So, we analyzed in the cases when $H(c_w) > 4$.
5. The *k*-means method is run 10 times, then averaged.

# References

Aloteibi, S., & Sanderson, M. (2014). Analyzing geographic query reformulation: An exploratory study. *J. Am. Soc. Sci. Technol.*, *65*(1), 13–24.

Bai, Y. B., Wu, S., Ren, Y., Ong, K., Retscher, G., Kealy, A., ... Zhang, K. (2014). A New Approach for Indoor Customer Tracking Based on a Single Wi-Fi Connection. In Ipin '14.

Bates, M. J. (2010). Information Behavior. In Encyclopedia of Library and Information Sciences, 3, 2381–2391.

Beitzel, S. M., Jensen, E. C., Chowdhury, A., Grossman, D., & Frieder, O. (2004). Hourly Analysis of a Very Large Topically Categorized Web Query Log. In Proceedings of the 27th annual international acm sigir conference on research and development in information retrieval (pp. 321–328). New York, New York, USA.

Bell, S., Jung, W. R., & Krishnakumar, V. (2010). Wifi-based enhanced positioning systems: Accuracy through mapping, calibration, and classification. In Isa '10 (pp. 3–9).

Biczok, G., Martinez, S., Jelle, T., & Krogstie, J. (2014). Navigating MazeMap: indoor human mobility, spatio-logical ties and future potential. CoRR.

Chua, A. Y. K., Balkunje, R. S., & Goh, D. H.-L. (2011). Fulfilling mobile information needs: A study on the use of mobile phones. In Icuimc '11.

Church, K., & Oliver, N. (2011). Understanding Mobile Web and Mobile Search Use in Today's Dynamic Mobile Landscape. In Mobilehci'11 (pp. 67–76).

Church, K., & Smyth, B. (2009). Understanding the intent behind mobile information needs. In Iui '09.

Church, K., Smyth, B., Cotter, P., & Bradley, K. (2007, May). Mobile information access: A study of emerging search behavior on the mobile Internet. ACM Trans. Web, 1(1).

Cui, Y., & Roto, V. (2008). How people use the web on mobile devices. In Www '08 (pp. 905–914).

Davies, D. L., & Bouldin, D. W. (1979). A Cluster Separation Measure. IEEE Trans. Pattern Anal. Mach. Intell., 1(2), 224–227.

De Domenico, M., Lima, A., & Musolesi, M. (2013, December). Interdependence and predictability of human mobility and social interactions. Pervasive Mob. Comput., 9 (6), 798–807.

Duarte Torres, S., Hiemstra, D., & Serdyukov, P. (2010). Query Log Analysis in the Context of Information Retrieval for Children. In Proceedings of the 33rd international acm sigir conference on research and development in information retrieval (pp. 847–848). New York, New York, USA: ACM Press.

Evans, K. R., Christiansen, T., & Gill, J. D. (1996). The impact of social influence and role expectations on shopping center patronage intentions. Journal of the Academy of Marketing Science, 24(3), 208–218.

Gan, Q., Attenberg, J., Markowetz, A., & Suel, T. (2008). Analysis of Geographic Queries in a Search Engine Log. In Proceedings of the 1st international workshop on location and the web (pp. 49–56). Beijing, China: ACM.

Hinze, A. M., Chang, C., & Nichols, D. M. (2010). Contextual queries express mobile informa- tion needs. In Mobilehci '10 (pp. 327–336).

Hodkinson, C., Kiel, G., & McColl-Kennedy, J. R. (2000, May). Consumer web search behaviour: Diagrammatic illustration of wayfinding on the web. Int. J. Hum.-Comput. Stud., 52 (5), 805– 830.

Jansen, B. J. (2000, March). Real life, real users, and real needs: a study and analysis of user queries on the web. Information Processing and Management, 36(2), 207–227.

Jansen, B. J., Ciamacca, C. C., & Spink, A. (2008, June). An Analysis of Travel In- formation Searching on the Web. Information Technology & Tourism, 10(2), 101–118.

Jensen, C. S., Lu, H., & Yang, B. (2010). Indoor-a new data management frontier. IEEE Data Engineering Bulletin, 33(2), 12–17.

Kamvar, M., & Baluja, S. (2006). A large scale study of wireless search behavior: Google mobile search. In Chi '06 (pp. 701–709).

Kang, H.-Y., Kim, J.-S., & Li, K.-J. (2009). Similarity measures for trajectory of moving objects in cellular space. In Sac '09 (pp. 1325–1330).

Khare, A. (2012, October). Influence of mall attributes and demographics on Indian consumersâĂŹ mall involvement behavior: An exploratory study. Journal of Targeting, Measurement and Analysis for Marketing, 20(3-4), 192–202.

Kjærgaard, M. B., Krarup, M. V., Stisen, A., Prentow, T. S., Blunck, H., Grønbæk, K., & Jensen, C. S. (2014). Indoor positioning using wi-fi–how well is the problem understood? In Ipin '13.

Kumar, R., & Tomkins, A. (2010). A Characterization of Online Browsing Behavior. In Www '10 (pp. 561–570).

Lee, I., Kim, J., & Kim, J. (2005). Use contexts for the mobile internet: A longitudinal study moni- toring actual use of mobile internet services. International Journal of Human-Computer Inter- action, 18 (3), 269-292.

Misra, A., & Balan, R. K. (2013, December). Livelabs: Initial reflections on building a large- scale mobile behavioral experimentation testbed. SIGMOBILE Mob. Comput. Commun. Rev., 17(4), 47–59.

Mobasher, B., Cooley, R., & Srivastava, J. (2000). Automatic personalization based on web usage mining. Communications of the ACM, 43(8), 142–151.

Nylander, S., Lundquist, T., & Brännström, A. (2009). At Home and with Computer Access âĂŞ Why and Where People Use Cell Phones to Access the Internet. In Chi (pp. 1639–1642).

Okabe, A., Boots, B., Sugihara, K., & Chiu, S. N. (1999). Spatial tesselations: Concepts and applications of voronoi diagrams (2nd ed.) [Book]. John Wiley and Sons.

Pandey, S., Punera, K., Fontoura, M., & Josifovski, V. (2010). Estimating Advertisability of Tail Queries for Sponsored Search. In Proceedings of the 33rd international acm sigir conference on research and development in information retrieval (pp. 563–570). New York, New York, USA: ACM Press.

Regalado, A. (2013). It's all e-commerce now [Electronic Book Section].

Ren, Y., Tomko, M., Ong, K., & Sanderson, M. (2014). How people use the web in large indoor spaces. In Cikm '14 (pp. 1879–1882).

Richter, K.-F., Winter, S., & Santosa, S. (2011). Hierarchical representations of indoor spaces. Environment and Planning-Part B, 38(6), 1052.

Ruetschi, U.-J. (2007). Wayfinding in scene space: Transfers in public transport. Phd. dissertation.

Sanderson, M., & Kohler, J. (2004). Analyzing geographic queries. In Proc. of the workshop on geographic information retrieval. Sheffield, UK.

Schulz, D., Bothe, S., & Körner, C. (2012). Human mobility from gsm data-a valid alternative to gps. In Mobile data challenge 2012 workshop, june (pp. 18–19).

Sen, S., Chakraborty, D., Subbaraju, V., Banerjee, D., Misra, A., Banerjee, N., & Mittal, S. (2014). Accommodating user diversity for in-store shopping behavior recognition. In Iswc '14 (pp. 11–14).

Silverstein, C., Marais, H., Henzinger, M., & Moricz, M. (1999). Analysis of a Very Large Web Search Engine Query Log. In Acm sigir forum (pp. 6–12).

Sohn, T., Li, K. A., Griswold, W. G., & Hollan, J. D. (2008). A diary study of mobile information needs. In Chi '08 (pp. 433–442).

Song, Y., Ma, H., Wang, H., & Wang, K. (2013). Exploring and exploiting user search behavior on mobile and tablet devices to improve search relevance. In Www '13 (pp. 1201–1212).

Spink, A., Jansen, B., Wolfram, D., & Saracevic, T. (2002). From e-sex to e-commerce: Web search changes. Computer, 35(3), 107–109.

Spink, A., Wolfram, D., Jansen, M. B., & Saracevic, T. (2001). Searching the Web: The Public and Their Queries. J. Am. Soc. Inf. Sci. Technol., 53(3), 226–234.

Teevan, J., Karlson, A., Amini, S., Brush, A. J. B., & Krumm, J. (2011). Understanding the importance of location, time, and people in mobile local search behavior. In Mobilehci '11 (pp. 77–80).

Vernor, J. D., Amundson, M. F., Johnson, J. A., & Rabianski, J. S. (2009). Shopping Center Ap- praisal and Analysis.

Wan-chik, R., Clough, P., & Sanderson, M. (2013). Investigating religious information searching through analysis of a search engine log. J. Am. Soc. Inf. Sci. Technol., 64(12), 2492–2506. doi: 10.1002/asi

West, R., White, R. W., & Horvitz, E. (2013). From cookies to cooks: Insights on dietary patterns via analysis of web usage logs. In Www '13 (pp. 1399–1410).

Wiener, J. M., Büchner, S. J., & Hölscher, C. (2009). Taxonomy of human wayfinding tasks: A knowledge-based approach [Journal Article]. Spatial Cognition & Computation, 9(2), 152-165.

Wolfram, D. (2008, May). Search characteristics in different types of Web-based IR environments: Are they the same? Information Processing & Management , 44 (3), 1279–1292.

Wolfram, D., Spink, A., Jansen, B. J., & Saracevic, T. (2001). Vox populi: The public searching of the web. J. Am. Soc. Inf. Sci. Technol., 52(12), 1073–1074.